\documentclass[12pt]{article}%
\usepackage{mathpazo}%
\usepackage{pifont,latexsym}%
\usepackage[fleqn]{amsmath} %
\usepackage{amssymb,euscript} %
\usepackage{float,endnotes}%
\usepackage[para]{footmisc}%
\usepackage{tocloft}%
\usepackage{xcolor}
\newcommand{\ncom}{\newcommand}%
\newcommand{\rcom}{\renewcommand}%
\ncom{\mrm}{\mathrm}%
\ncom{\tsf}[1]{\mbox{\rm \textsf{#1}}}%
\ncom{\trm}{\textrm} %
\ncom{\tbf}{\textbf} %
\ncom{\sfS}{\tsf{S}} %
\ncom{\sfM}{\tsf{M}} %
\ncom{\eulA}{\EuScript{A}} %
\ncom{\eulB}{\EuScript{B}} %
\ncom{\eulW}{\EuScript{W}} %
\ncom{\wP}{\widehat{P}}%
\ncom{\wQ}{\widehat{Q}}%
\ncom{\bm}[1]{\mbox{\boldmath{$#1$}}}%
\ncom{\ensmath}{\ensuremath}%
\ncom{\R}{\ensmath{\mathbb{R}}}%
\ncom{\N}{\ensmath{\mathbb{N}}}%
\ncom{\C}{\ensmath{\mathbb{C}}}%
\ncom{\Q}{\ensmath{\mathbb{Q}}}%
\ncom{\V}{\ensmath{\mathbb{V}}} %
\ncom{\rmi}{\trm{i}}%
\ncom{\la}{\langle}%
\ncom{\ra}{\rangle}%
\ncom{\bla}{\big\la}%
\ncom{\bra}{\big\ra}%
\ncom{\calH}{\mathcal{H}}%
\ncom{\calI}{\mathcal{I}}%
\ncom{\calP}{\mathcal{P}}%
\ncom{\bfQ}{\textbf{Q}}%
\ncom{\bfq}{\textbf{q}}%
\ncom{\bfp}{\textbf{p}}%
\ncom{\fnsize}{\footnotesize}%
\rcom{\emptyset}{\varnothing}%
\ncom{\beq}{\begin{equation}}%
\ncom{\enq}{\end{equation}}%
\ncom{\terug}{{\protect\hspace*{-2ex}}} %
\ncom{\blsq}{\ensuremath{\blacksquare}} %
\rcom{\to}{\rightarrow}%
\ncom{\mbbm}[1]{\mbox{\boldmath{$#1$}}}%
\ncom{\bfell}{\mbbm{\ell}} %
\ncom{\conj}{\mbbm{\;\wedge\;}}%
\ncom{\disj}{\mbbm{\;\vee\;}}%
\ncom{\negate}{\raisebox{0.2ex}{\boldmath{$\neg$}}}%
\ncom{\ifthen}{\mbbm{\;\longto\;}}%
\ncom{\ifff}{\mbbm{\;\longleftrightarrow\;}}%
\ncom{\entails}{\mbbm{\vdash}}%
\ncom{\all}{\mbbm{\forall}\,}%
\ncom{\eris}{\mbbm{\exists}\,}%
\ncom{\up}{\protect\vspace*{-1ex}} %
\ncom{\bskip}{\mbox{}\protect\\[0.5em]\noindent}%
\rcom{\baselinestretch}{1.20}%
\ncom{\regelwit}{\mbox{}\newline\newline}%
\ncom{\minqm}{\ensmath{\mbox{\small QM}_{0}}}%
\ncom{\stqm}{\ensmath{\mbox{\small OxQM}}}%
\ncom{\evqm}{\ensmath{\mbox{\small EvQM}}}%
\ncom{\copqm}{\ensmath{\mbox{\small CopQM}}}%
\ncom{\bqm}{\ensmath{\mbox{\small BQM}}}%
\ncom{\citaat}[1]{\\[0.8em]\indent{\parbox[t]{0.96\textwidth}%
   {\small {#1}}}\\[1.2em]}%
\ncom{\display}[1]{\\[0.8em]\indent{\parbox[t]{0.96\textwidth}%
   {#1}}\\[1.2em]}%
\newenvironment{famlist}%
{\begin{list}{}%
  {\setlength{\labelwidth}{-\parindent} %
  \setlength{\labelsep}{1.5ex} %
  \setlength{\leftmargin}{\parindent} %
  \setlength{\itemindent}{-\labelsep}%
  \setlength{\itemsep}{-0.5ex}%
   } }%
{\end{list}}
\newenvironment{famlistdik}%
{\begin{list}{}%
  {\setlength{\labelwidth}{-\parindent} %
  \setlength{\labelsep}{1.5ex} %
  \setlength{\leftmargin}{\parindent} %
  \setlength{\itemindent}{-\labelsep}%
  \setlength{\itemsep}{0ex}%
   } }%
{\end{list}}
\newenvironment{refs}%
{\begin{list}{}%
 {\setlength{\labelwidth}{0pt} %
  \setlength{\leftmargin}{\parindent} %
  \setlength{\itemindent}{-1.0\parindent}%
  \setlength{\itemsep}{-0.5ex}%
   } }%
{\end{list}}
\rcom{\contentsname}{\normalsize Table of Contents}%
\rcom{\cftsecfont}{\small }%
\rcom{\cftsubsecfont}{\small }%
\rcom{\cftdotsep}{6} %
\rcom{\cftsecleader}{\cftdotfill{\cftdotsep}}%
\rcom{\cftsecpagefont}{\normalfont\small } %
\rcom{\cftaftertoctitleskip}{-15pt} %
\rcom{\cftsetrmarg}{2em} %
\rcom{\cftsetpnumwidth}{1em} %
\begin{document}%
\raggedbottom%
\thispagestyle{empty}
\noindent%
\begin{flushright}
{\scriptsize \fbox{For a Festschrift for P.C.\ Suppes, celebration of his 90th birthday in 2012, 
to appear in some \emph{Festschrift}}}
\end{flushright}
\mbox{}\\[2em]
{\LARGE \textbf{Circumveiloped by Obscuritads.}}\\[1ex]
{\large \textbf{The nature of interpretation in quantum mechanics,\\[0.5ex]
hermeneutic circles and physical reality, with cameos\\[0.5ex]
of James Joyce and  Jacques Derrida}}\endnote{The main title of this paper is 
borrowed from James Joyce's \emph{Finnegans Wake} (1939, 244.15), as  are the
titles of the Sections, for reasons that will become evident to the imaginative mind
as we proceed; the notation `244.15' is standard and means: page 244, line 15. Any edition
can be consulted, because they all use the same pagination and lining.}\\[2em]
{\Large \emph{\textbf{F.A.\ Muller}}}\\[2ex]
\emph{31 December 2013}
\begin{quote}
{\small \textbf{Summary.}~~The quest for finding 
the \emph{right} interpretation of Quantum Mechanics ({\fnsize QM})
is as old als {\fnsize QM} and still has not ended, and may never end.
The question \emph{what an interpretation of {\fnsize QM} is} 
has hardly ever been raised explicitly, let alone answered. We raise it and answer it.
Then the quest for the right interpretation can continue \emph{self-consciously},
for we then know \emph{exactly} what we are after.
We present a list of minimal requirements that something
has to meet in order to qualify as \emph{an interpretation of {\fnsize QM}}.
We also raise, as a side issue, the question how the discourse on the interpretation 
of {\fnsize QM} relates to hermeneutics in Continental Philosophy.\\
\mbox{}}
\tableofcontents
\end{quote}
\thispagestyle{empty}
\clearpage
\setcounter{page}{1}
\section{Nuemaid Motts and a Nichtian Glossary}
{\up}James Augustine Aloysius Joyce (1882--1941)
constructed this tantaltuous and tumulising towertome \emph{Finnegans Wake} (1939)
in the period during which quantum mechanics was \emph{created} (1923--1939),
by Werner Heisenberg, Max Born, Pascual Jordan, Wolfgang Pauli,
P.A.M.\ Dirac and Erwin Schr\"{o}dinger, was \emph{axiomatised}, by Johnny
von$\,$Neumann, was \emph{applied}, by numerous physicists, 
was \emph{interpreted}, by Niels Bohr 
and Heisenberg, was \emph{demonstrated to exclude} certain alternative 
theories, by Von$\,$Neumann, and was \emph{criticised}, by Albert Einstein, 
Schr\"{o}dinger and others.
Over the past decades, philosophers have joined the interpretation effort ---
with remarkable success, we dare add.

News broadcast in \emph{Finnegans Wake}:\endnote{Joyce [1938],  353.22--23.}
\citaat{The abnihilization of the etym by the grisning of the grosning of
the grinder of the grunder by the first lord of Hurteford expolodonates
through Parsuralia with an ivanmorinthorrorumble fragoromboassity amidwhiches general uttermost
confusion are perceivable moletons scaping with mulicules $\ldots$ Similar scenatas are 
projectilised from Hullulullu, Bawlawayo, empyreal Raum and mordern Atems.}
Recall that in 1911 Lord Rutherford (lord of Hurteford)
split the atom (etym), a detonation of sorts where electrons
(moletons) and molecules (mulicules) escape, 
projectiles moving through empirical space (German
\emph{Raum}). The historical event was reported all around the globe,
like in Paris (Hullulullu), Rome (Bawlawayo) 
and Athens (Atems). 

Traces of both the Quantum and the Relativity Revolution
in physics are scattered all over \emph{Finnegans Wake}.

Philosophically, \emph{Finnegans Wake} can be seen to raise the issue of what 
\emph{meaning} is, even of \emph{language} is. We let this grand issue sleep.

\emph{To interpret} a word, an expression, a sentence, 
a text, is \emph{to assign meaning to} it.
Clear and unambiguous kinds of texts, such as the telephone 
directory of Hullulullu, the weather forecast for tomorrow in Bawlawayo,
the papers of Patrick Colonel Suppes,
and the user manual of your brand new ten-dimensional retina-screen nanowave 
stringphone, do not stand in need of
interpretation. Other kinds of text cry out earsplittingly for interpretation, of which 
\emph{Finnegans Wake} arguably is the most clear and unambiguous
instance ever created. Quantum mechanics is somewhere in between. As 
Dummett (1925--2011) testified:\endnote{Dummett  [1991], p.~13.}
\citaat{Physicists know how to use quantum mechanics and,
impressed by its success, think it is \emph{true}; but their endless
debates about the interpretation of quantum mechanics show that
they do not know what it \emph{means}.}
\indent But standing in need of interpretation is something that
\emph{Finnegans Wake} and quantum mechanics ({\small QM}) share with 
lots of other texts. We must take a closer look to understand why
they are special.

Joyce judged \emph{natural-language-as-we-know-it} (whenceforth: Nalasweknowit)
inadequate to describe what happens in the dream world, and created,
for this very purpose, a `new language',
\emph{if} that is the appropriate phrase: 
``nuemaid motts truly plural and plusible'' 
(138.08--09) and a ``nichtian glossery which purveys aprioric roots for 
aposteriorious tongues this is nat language in any sinse of the world'' (83.10--11).
Heisenberg and Bohr judged Nalasweknowit, of 
which they considered  `the language of classical physics' a refinement, 
inadequate to describe what happens in the microphysical
world, the world of \emph{very} small physical entities and \emph{very} brief
physical processes. 

Small wonder. Nalasweknowit has developed
while \emph{homo sapiens} and its ancestry was wide awake, i.e.\ not dreaming, 
and interacting with 
the macrophysical world filled with trees, rocks and animals, and with days, seasons
and lifetimes. Man was occupied with fulfilling his biological needs of nutrition, 
protection and procreation, rather than with penetrating the ephemerally 
flashing realm of dreams, explaining the phenomena by means of theories, 
or unravelling the mysteries of a realm of reality inaccessible by the unaided senses. No one 
had ever wanted or needed to go above and beyond the waking macrophysical world, or
to transcend our biological needs, which we shared with the beasts. 
But, at some day, the time had come that we did want and did need to go 
precisely there, and we did want transcend our beasty needs. On Earth, \emph{how}?

Back to the early 20th Century.
Understanding the microphysical world was no longer deemed possible with 
Nalasweknowit. In order to grasp this realm of reality somehow, only a `symbolic description', or a \emph{Deutung}, by abstract mathematical means seemed possible. 
Of course nothing remotely like ``multimathematical immaterialities''
(394.31--32) was the means for Joyce to penetrate the realm of dreams.
Joyce constructed numerous neologisms and \emph{portmanteaux}: 
``the dialytically separated elements of precedent decomposition
for the verypetpurpose of subsequent recombination'' (614.34--35).
In contradistinction to how Joyce accomplished his daunting task of evoking
the phantasmagorical events of deep weep sleep, i.e.\ by creating \emph{Finnegans Wake},
and thereby replacing Nalasweknowit, what the founding fathers of {\small QM}
did was far less radical: a comparatively small yet
significant enrichment of Nalasweknowit would initially turn out to be sufficient
to unlock the secrets of the atom --- but would eventually 
also lead to perplexities the world of science had never seen before $\ldots$
\section{Abnihilazation and Everintermutuoemergent}
{\up}No matter how one characterises {\small QM} pecisely, e.g.\ as the deductive
closure of a set of sentences (the postulates) in a formal language or through a class
of models (structures in the domain of discourse of axiomatic set-theory), 
or some sophisticated combination of these,
{\small QM} incontestably has \emph{propositional content},
expressed in declarative sentences of Nalasweknowit,
enriched with physical and mathematical vocabulary and with symbols. {\small QM}
makes a large variety of pronouncements about physical reality, measurements included, 
that can be and have been tested severely. 
Sometimes {\small QM} says things that raise our eyebrows sky high, 
like there be non-local correlations that do \emph{not} fall off 
with distance and \emph{cannot be explained} even by an appeal to the entire past of the 
carriers of the correlata (version of Bell's Theorem), and like
a continuously \emph{observed} kettle filled with water on the fire that never boils 
(quantum Zeno paradox).\endnote{For the sake of clarity: we suppose that to observe is to measure, so
by contraposition, not to measure is not to observe; to measure is not necessarily to observe.
This is correct, because think of, say, measuring the presence of a neutrino or the energy of
an electron, which are unobservable entities: we measure but cannot observe.}
Sometimes {\small QM} remains mute when we desperately crave for answers, like when we ask whether Schr\"{o}dinger's unmeasured, and therefore unobserved
cat \emph{is} dead or alive, since {\small QM} does neither fulfil the 
truth-condition for the sentence `The \emph{unobserved}
cat is alive', nor for `The \emph{unobserved} cat is dead',  {\small QM}
falls silent. Needless to add that the celebrated case of Schr\"{o}dinger's cat 
extrapolates to the entire unmeasured part of the universe, which
comprises nearly everything. We observe a few drops of the ocean of being. Nearly all of 
physical reality is \emph{ontically indeterminate}, and therefore is not \emph{really} 
reality at all $\ldots$  {\small QM} forbids us to speak whereof we want to speak.

Notice that a use theory of meaning, which takes the use of words, expressions and sentences
constitutive for their meaning, does not sit comfortably either with Dummett's locution displayed above: if, \emph{first}, knowing the meaning of {\small QM} resides in knowing how to use it, and, \emph{secondly}, granted that physicists know how to use {\small QM} in every which way, that is,
knowing how to construct quantum-mechanical models of phenomena, knowing how to reason
quantum-mechanically, knowing how to calculate measurement outcomes, knowing how analyse
experiments using {\small QM} then they should \emph{know} its meaning, 
whereas the endless debates about the interpretation of {\small QM} --- which we shall provisionally call 
its \emph{hermeneutic predicament} --- is taken to show the 
contrary, namely that they \emph{do not know} what {\small QM} means.

If the project \emph{to interpret} {\small QM} is, in good hermeneutic fashion,
to assign meaning to it, we must ask which expressions of {\small QM}
stand in need of interpretation, because, then, apparently \emph{their}
meaning is not obvious, or is ambiguous,
or is obscure, or in any way stand in dire need of receiving clear and unambiguous
meaning. If every expression in {\small QM} were perfectly clear, there would 
obviously be no need to interpret {\small QM}.

The vocabulary of {\small QM} is rather mathematical and its
mathematical concepts are crystal clear. They do not stand in need of interpretation
--- Hilbert-space, self-adjoint operator, eigenvalue equation, unitary evolution,
statistical operator, Clebsch-Gordan coefficients, Weyl rays,
unitary representations of a symmetry group, permutation operators,
Wigner distributions, and what have you. The physical vocabulary, 
including physical magnitude, physical system,
composite system and subsystem, physical property and physical relation
also seem far from obscure. This is not to say that these
concepts are beyond interpretation, let alone beyond metaphysical disputation. 
Concepts of {\small QM} that stand 
in need of interpretation are the physical state of a physical system and the probability
for finding specific outcomes upon measurement, and certainly
the concept of measurement itself. On the one hand, one can send everybody who 
raises questions about measurement to a laboratory: observe what is happening there and
ask around; if that will not do, then nothing will. On the other hand, when we ask
what a measurement is, we are after a general answer, a general concept of measurement,
one that encompasses what happens inside all laboratories; 
everything we want to call a measurement should be an instance of our general concept,
and everything we do not want to call a measurement should not be an instance of it. 
This general
concept should cover our use of the word `measurement', but need not cover it entirely,
for we shall gladly pay the price of lack of full coverage for a clear general concept.
In short, we are after a Carnapian explication of the concept of measurement. 
By way of an interjection, we shall have a go at this in the next Section.
When a physical system qualifies as piece of measurement apparatus, when a 
physical interaction qualifies as a measurement interaction, when an event 
qualifies as a measurement event,
and perhaps more, have been issues for analysis and controversy since the
advent of {\small QM}. Certainly we want to count these issues part and parcel
of discourse `the interpretation of {\small QM}'. 

\emph{Probability} is mathematically represented by a normed
additive mapping from some Boolean subset family of $\R$, say
the intervals $\calI(\R)$, to the interval $[0,1]\subset\R$:
\beq \textrm{Pr}:~\calI(\R)\to\R\;.\enq
So for the mathematician, this is all there is to probability: a 
normed measure on $\calI(\R)$. 
Not for the scientist, who has to relate the normed measure
to the world. The quantum-mechanic has to relate probability at the very least
to measurement outcomes. The only way to do this is via relative frequencies. But whether 
probability \emph{is} a limiting relative frequency amounts to taking a
further philosophical step, as does identifying probability with \emph{objective
change}, as does identifying it with \emph{propensity}, i.e.\ some \emph{generalised quantitative
disposition}, and as does taking it as a \emph{degree of subjective belief} or a 
\emph{degree of rational credence}. We entered the field of interpreting
probability. Some hold that \emph{quantum probability} is somehow special
and different from probability as it occurs elsewhere in physics and in 
science generally. The discussion of whether this is true, and how quantum
probability then differs from probability applied elsewhere is however
not a central theme in the interpretation of {\small QM} --- at best it is a
peripheral theme. 

Like the concept of probability, the concept of \emph{physical state} is primitive,
yet unlike probability, it can be and is represented mathematically 
in many distinct ways: as a
\ncom{\blast}{$\circledast$}
\begin{famlist}
\setlength{\itemsep}{-1ex}
{\item[\blast] a normed Hilbert-vector, or}
{\item[\blast] a Weyl-ray, or}
{\item[\blast] a statistical operator acting on a Hilbert-space, or}
{\item[\blast] a positive map on a $C^{\ast}$-algebra.}
\end{famlist}
Maybe calling a Hilbert-vector (or Weyl-ray, or $\ldots$) \emph{the mathematical 
representative of the physical state}
of a physical system is a \emph{mistake}: a Hilbert-vector should remain  a physically 
uninterpreted and purely mathematical concept in {\small QM}, an auxiliary device to calculate 
probability distributions of measurement outcomes. There is no `physical 
state' of the unmeasured cat in purgatory:
we are led to believe that the cat has, or is in, a \emph{physical state} 
by mistakenly trying to attribute physical meaning to
a Hilbert-vector that is a superposition of two vectors, which according to the standard
property postulate we associate with a 
cat having the property of being dead and one having the property of being alive, respectively. 
We believe the unmeasured cat is some particular physical
state but perhaps it isn't. {\small QM} associates a Hilbert-vector to the cat, which is devoid
of physical meaning, but enables the computation of probability measures over 
measurement-outcomes,
which are full of physical meaning. Thus we have physical meaningfulness out of physical 
meaninglessness. Sheer magic. Magic does however not help us to understand 
physical reality.

The wilful jump to meaninglessness seems however a cheap way out. I don't like it.
We believe that the unmeasured cat is either stone dead or breathing, because
\emph{tertium non possibilium},
and we want {\small QM} to be logically compatible with this belief, 
at the very least, and preferrably to imply one or the other belief.
After all, {\small QM} also predicts that as soon as we peek at (i.e.\ measure) 
the cat, through a
pinhole, unbeknownst to the cat, it \emph{is} either dead or alive. Rather than to
withhold physical significance from the Hilbert-vector, 
we should try to assign physical significance to it (or to a Weyl-ray, or $\ldots$).
For how else could it determine physically meaningful probability measures over
measurement-outcomes? No physical significance in, but physical significance out?
That ought to be unacceptable.
One way is to connect Hilbert-vectors to equivalence classes of preparation
procedures in the laboratory. This won't help us however with
Schr\"{o}dinger's unmeasured cat. This won't help us with anything, because
superpositions are the rule, not the exception. The founding fathers of 
{\small QM} started with electrons in superpositions, soon other elementary
particles followed, then atoms, and nowadays we have bucky-ball molecules
and circulating currents in superconducting metals in superpositions in the laboratory. 
The march of superpositions from the realm of the tiny to the realm of medium-sized
dry objects is not halting.

So-called \emph{modal interpretations} of {\small QM} have taught us 
that the cat ceases to be
a problem as soon as we reject `half' of what we shall call the
\emph{Standard Property Postulate} of {\small QM},
which one can find the classic texts of  Von$\,$Neumann [1932] and Dirac [1928] 
--- and which remains nearly always tacit in textbooks on {\small QM}.\endnote{Any
author on QM who presents Schr\"{o}dinger's cat as a problem in that it is 
neither dead nor alive, tacitly assumes that it is necessary for the cat to be in a relevant eigenstate in order to be either dead or alive. The Standard Property Postulate is also known as `the eigenstate-eigenvalue link'.}
\display{\blsq~\textbf{Standard Property Postulate (Dirac, Von$\;$Neumann)}. 
\emph{A physical  system $\sfS$ 
having physical state $|\psi\ra\in\calH$ has quantitative physical property
mathematically represented by the ordered pair $\la B,b\ra$,
where $B$ is an operator representing some physical magnitude and where $b\in\R$,
iff $\;|\psi\ra$ is an eigenstate of $B$ having eigenvalue $b$}:
$B|\psi\ra=b|\psi\ra$.\endnote{Extensions to mixed states are possible.}}
\indent When it is no longer necessary for the state to be an eigenstate of $B$ in order
for physical system $\sfS$ to have a property of the sort $\la B,b\ra$, then the
unmeasured cat \emph{can} be
either dead or alive even when its state is \emph{not} a corresponding eigenstate ---
but is a superposition of such eigenstates.
The compatibility between {\small QM} and our belief that the unmeasured cat is either dead
or alive is saved. What can be adhered to, then, is not the Standard Property Postulate
but the \blsq~\textbf{Sufficiency Property Postulate}, according to which it
is sufficient (but not necessary) for the system to be in some eigenstate of $B$ in order to
possess property $\la B,b\ra$ (one drops one conjunct of the Standard Property Postulate).

Logically weakening a postulate seems however to have little to do
with \emph{interpretation} in the hermeneutic sense of assigning meaning to expressions
whose meaning is unclear, ambiguous or obscure. Indeed, for modal interpreters
of {\small QM}, the problem of interpretation is to find \emph{the right conditions
for property ascriptions} --- in addition to the stingy Sufficiency Property Postulate ---, 
rather than to dwell on the meaning of `physical state'
(or Weyl-ray, or $\ldots$). (We say `stingy', because a physical system is almost
never in an eigenstate, so one can almost never invoke the Sufficiency Property Postulate.)
This points away from hermeneutical activity when considering interpreting {\small QM}
to changing the postulates --- unless one subscribes to a theory of meaning such
that changing the conditions for the ascription of properties changes the meaning of
the word `property', in which case one should consider such property postulates
as Carnapian \emph{meaning postulates}, rather than synthetic postulates that
are made true (or false) by the way the world is.

It is in order to mention the exception of Oxonian Everettians, who under the
lead of S.W.\ Saunders tinker with the meaning of `existence' and tensed expressions
by relativising them to a `perspective', a `branch', and who, like all Everettians,
assign special significance to the terms of the state vector when expanded in
a special basis, which is selected by the physical proces of decoherence.\endnote{See
Wallace [2013] for a state of the art defence of the Everett Interpretation.}
They re-interpret and therefore change the meaning of words in Nalasweknowit. Hermeneutics in action.
One could also maintain that the problem of interpreting {\small QM} just
is the problem of finding an intelligible physical meaning to attribute to the mathematical concept of a Hilbert-vector (or $\ldots$) in such a way that our belief that
the unmeasured cat is either dead or alive survives whilst leaving the Von$\,$Neumann
postulates of {\small QM} untouched in all their glory, save perhaps minor modifications.
But then modal interpreters of {\small QM} are \emph{not interpreting} {\small QM}. 
There is no hermeneutic activity going on. What, then, \emph{are} they doing?

They are changing the theory of {\small QM} by \emph{changing} (one of) 
its postulates, which results in a \emph{different} theory of {\small QM},
just like changing the parallel axiom of Euclidean Geometry 
results in a \emph{different} geometrical theory. When that 
different geometrical theory, if true, tells us that the structure of space
is different from what Euclidean Geometry tells us, then \emph{mutatis mutandis}
modal {\small QM} provides a different description of the microphysical world 
than standard {\small QM} does. This is the key insight of this paper and the essence
of our alternative view of what it means \emph{to interpret} {\small QM}. But before we
turn to that, first the promised interjection on measurement.
\section{Multimathematical Murkblankered Immaterialities}
{\up} \subsection{Preambule}
{\up}In English, as in most languages, \emph{to measure} is a \emph{verb}.
The \emph{noun}  `measurement' is derived from it: to measure is to perform
a measurement, and to perform a measurement is to measure. 
To measure is a manifestation of intentional behaviour,
i.e.\ it is a type of \emph{action}, performed by a human being, with a purpose --- 
or by any being having the cognitive capacities to exhibit it. 
Therefore the concept of measurement is an \emph{intentional} concept. 

The concept of measurement is expressed most explicitly, we submit, by a pentatic predicate:
\emph{someone} ($p$) \tsf{measures} \emph{something} ($\eulA$) 
that pertains to \emph{something} ($\sfS$)
using \emph{something} else ($\sfM$) and obtains \emph{result} $a$:
\beq \trm{Measure($p,\,\eulA,\,\sfS,\,\sfM,\,a$)}:~~
\trm{$p$ measures $\eulA$ of $\sfS$ by means of $\sfM$ and obtains $a$}.
\label{Meas}\enq
\indent There are \emph{kinds} of measurements, whose extensions are subclasses
of the extension of \eqref{Meas}:
demolishion measurements, ideal measurements, extensive measurements,
perfect measurements, sharp measurements, weak measurements, $\ldots$
The word \tsf{measurement} occurs in combinations with other words, 
especially in science; these combinations express different but allied concepts, which
we call \emph{measurement concepts}: measurement event, measurement process,
measurement procedure, measurement result, outcome,
measurement interaction, measurement apparatus,
measurement theory, measurability. In every case, the suffix `measurement' points to a
\emph{kind}: measurement events are a \emph{kind} of events, they form a 
subclass of the class of all events; measurement processes are a \emph{kind} of processes, they form a subclass of the class of all processes; \emph{etc}. 
The purpose of this Section is to analyse the concept of measurement \eqref{Meas}
and other measurement concepts, but only those in so far needed in
the core concept Meas~\eqref{Meas}. The other measurement concepts will have to wait.

In our \emph{analysandum}, the concept of Measurement \eqref{Meas}, five things are connected:  human being $p$, value $a$, entity $\sfS$, entity $\sfM$, and
magnitude $\eulA$. The challenge is to characterise these concepts in a way that does not rely on the concept of measurement or any of the allied concepts, otherwise we awaken
the spectre of circularity. We gloss over the concept of a human being and
move now to the other concepts from the putative \emph{analysans}, one per
Subsection.
\subsection{Values}
\label{SubsValue}
{\up}Value $a$ is a number. Number $a$ is
a rational number ($a\in\Q$), because every measurement has a finite accuracy. 
Since two measurement results, $a$ and $b$, can
be taken as the real and the imaginary part of a complex number, there is room
for extending $\Q$ to $\C_{\trm{rat}}\subset\C$, the set of complex numbers having
rational real and imaginary parts. Nonetheless we continue with $a\in\Q$ 
and bracket $\C_{\trm{rat}}$.

To count is also a form of measurement, with a natural number as the result.
One can count the number of children in the class room with infinite accuracy:
there are $23$ children in the class room, or $23,000\ldots$,
and not $23\pm 1$, let alone $23,0\pm0,2$. (In these cases,
the outcome still is a rational number, because $\N\subset\Q$; so we can
stick to $a\in\Q$.)
\subsection{Entities}
\label{SubsEntityS}
{\up}We measure the emission spectrum \emph{of} Hydrogen; we measure the mass \emph{of} the Earth or \emph{of} a positron;
we measure the intensity \emph{of} the radio-active radiation \emph{of}
the nuclear power plant in Harrisburg; we measure the acidity \emph{of}
the liquid in this flask; \emph{etc.}
Clearly \emph{what} we measure, $\eulA$, always pertains to something ($\sfS$), 
and that something, that entity, we take to be a \emph{physical system}, as broadly construed  as possible: it consists of matter and fields, and is located in space-time.
This makes physical systems, in metaphysical parlance, \emph{concrete} rather than abstract
entities.
\subsection{Measurement Apparatus}
\label{SubsEntityM}
{\up}A measurement apparatus also is a physical system, that much seems clear.
We thus need a criterion to tell us \emph{which} physical systems qualify as 
a measurement apparatus and which do not. We proceed stepwise, (A)--(C): in each
step we consider a concept that we shall use in characterising what a measurement
apparatus is.

(A) \emph{Observability}. 
Surely a measurement apparatus $\sfM$ is a physical system that we,
human beings that measure, should be able to see (or hear $\ldots$). Otherwise $\sfM$ is
of no use to us! So $\sfM$ has to be \emph{observable} by us. 
This raises immediately the further question which physical systems are \emph{observable}.
Philosophers of science have pondered this question. We shall not repeat the ensuing literature but mention the rather obvious philosophical criterion for the extrinsic
property of observability. Let $p$ be a normal person, of sound mind and having
normal eye-sight. 
\begin{itemize}
{\item[]\tbf{Criterion for Observability.} Physical system $\sfS$ is \emph{observable} 
$\;$iff$\;$  for every $p$:  if $p$ were in front of $\sfS$ in broad daylight with open eyes, then 
$p$ would see $\sfS$.} 
\end{itemize}
Van Fraassen famously insisted that the observability of objects, events, facts, processes,
is a subject for scientific research, not for philosophical analysis. For a scientific characterisation of observability, see Muller [2005].

If $\sfS$ is observable, then $\sfS$ seems to have \emph{properties} that are observable, 
notably its shape and colours. What \emph{is} it that we actually \emph{see}? 
In full generality, this is a metaphysical question, which we wish to bracket.
We therefore limit ourselves to a characterisation of an observation predicate,
remaining neutral about whether predicates express universals or tropes. 
\begin{itemize}
{\item[]\tbf{Criterion for an Observation Predicate.} 
A predicate $F$ applied to physical system $\sfS$ is an \emph{observation predicate} $\;$iff
$\;$ for every $p$:  if $p$ were in front of $\sfS$ in broad daylight with open eyes, then $p$ 
would judge that $F(\sfS)$ or judge that $\negate F(\sfS)$ relying only 
on linguistic knowledge and on looking at~$\sfS$.}
\end{itemize}
\indent The addition of `relying only on linguistic knowledge' is to prevent that 
\emph{theory}, broadly construed, is relied on in order to judge whether $F(\sfS)$
or that $\negate F(\sfS)$. Suppose
around $2,000$~BC an Egyptian girl is taught that what we call `the sun'
is the god Raa. The other morning the girl wakes up, looks at the sky
and says: `A god has appeared in the sky.' She formed this judgement by
looking at the sky. But `being a god' should not qualify as an observation predicate.
It doesn't according to our Criterion, because `a god' relies on some
religious theory, that informs us about gods in general and Raa in particular.
That goes above and beyond \emph{linguistic knowledge}, i.e.\ knowledge of meaning,
knowledge of how to use words, the capacity to display appropriate linguistic behaviour by
uttering words, expressions and sentences in given circumstances, and
by understanding words, expressions and sentences when others utter them.  
Person $p$ must have some linguistic knowledge,
by the way, otherwise $p$ could not form the judgement that $F(\sfS)$ or that
$\negate F(\sfS)$. As soon as theoretical knowledge is needed to understand
what predicate $F$ means, $F$ cannot be an observation predicate.\endnote{Caveat:
the distinction between linguistic knowledge and theoretical knowledge is not exactly 
unproblematic, and even controversial. Similarly for the one between observation and
theoretical predicates. What to do when rejects these distinctions? Nothing.
Read on. Simply cut away this distinction from our \emph{analysans} of
measurement.} 

So much for the observability of measurement apparatus $\sfM$.

(B) \emph{One-one Correspondence}. 
When we read that the pointer of an Volt-meter points to $22\;$V,
we ascribe the property of an electric potential difference to a circuit; 
when I read $86$~kg on the display of a scale
while standing on it, I conclude that my body has a mass of $86$~kg;
\emph{etc.} So what we need is a one-one correspondence between
observable properties of $\sfM$ and values of the magnitude $\eulA$ that
$\sfM$ is measuring. Or better, \emph{intervals of values} rather than \emph{values} 
because of the finite
measurement accuracy: result $I=1.04\pm 0.07$~mA describes an observable property
of an Am-meter that corresponds to an infinite set of electric current
values, namely interval $[0.97,\;1.11]$.

(C) \emph{Relevant Interaction.} So a measurement apparatus $\sfM$ 
of magnitude $\eulA$ is an observable
physical system that leads to a one-one correspondence between certain
sets of values of $\eulA$ and observable properties of $\sfM$?

Almost right. Dupe can assign a rational number to 
few solid objects lying on the table in front of him using pencil and paper:
Dupe looks at an object and writes down some arbitrary rational number.
Dupe claims to have measured the masses of these objects, because we
have a one-one correspondence between observable properties of the paper
(the ink spots on it that express rational numbers) and values of
the physical magnitude mass of the objects. Yet surely the pencil and paper do not qualify as a 
\emph{measurement apparatus that measures mass}. 
Pencil and paper can be used \emph{to report} measurement outcomes, but
they are not themselves pieces of mass-measurement apparatus. 
Furthermore, just \emph{writing down an arbitrary rational number}
with a pencil on a piece of paper \emph{is not measuring} anything. 
If a one-one correspondence were enough, then measurement results would be what 
we want them to be, would become wholly under our control, whereas a 
measurement outcome seems 
to be something that is entirely beyond our control, something that has nothing to do with
what we want. Particular
measurement outcomes may be the ones we want, hope, wish, expect or fear.
But \emph{which} outcomes we shall actually obtain when we measure is beyond our control
and indifferent to our needs, hopes, wishes, expectations and fears.

Perhaps we should require that the one-one correspondence \emph{must be the result of a 
particular physical interaction} between measured object $\sfS$ and measuring 
object $\sfM$.
Dupe's one-one correspondence was not due to an interaction between the objects on
his table and the paper. Which particular physical interaction? The physical interaction
that occurs in explaining how $\sfM$ works, specifically how the one-one
correspondence between (sets of) values of $\eulA$ and (observation) predicates
that apply to $\sfM$ comes about. Let us call that physical interaction 
$\eulA$-\emph{relevant} --- which thus partly is an epistemic concept.

We arrive at the following criteria.
\begin{itemize}
{\item[]\textbf{Criterion for an $\eulA$-Measurement Apparatus.}~~Physical system $\sfM$ is a \emph{measurement apparatus of physical magnitude $\eulA$}, or briefly,
an $\eulA$-\emph{measurement apparatus},~~iff\\[0.5ex]
(M1)~~$\sfM$ is observable;\\
(M2)~~there is a one-one correspondence between (observation) predicates $F$ which
apply to $\sfM$, and sets of values of $\eulA$; and \\
(M3)~~the correspondence of (M2) is the result of the $\eulA$-relevant physical interaction
between physical system $\sfS$, to which $\eulA$ pertains, and $\sfM$.}
{\item[]\textbf{Criterion for a Measurement Apparatus.}~~Physical system $\sfM$ is 
a \emph{measurement apparatus}~~iff~~there is some physical magnitude $\eulA$ such that 
$\sfM$ is an $\eulA$-measurement apparatus.}
\end{itemize}

The young tree in the park garden is a measurement apparatus of the dichotomic
physical magnitude `presence of wind' ($\eulW$): if it oscillates 
visibly, then $\eulW$ has value $1$ (presence of wind), and if it remains unmoved, then 
$\eulW$ has value $0$ (absence of wind). Conclusion:
a piece of measurement apparatus need not be a \emph{technological artifact},
designed and constructed by human beings.
Mother Nature produces pieces of measurement apparatus too,
unintendedly, which is why being a technological artifact for $\sfM$ is not part of the criterion
for a measurement apparatus. 
\subsection{Magnitudes}
\label{SubsMagnitude}
{\up}Etymologically the word `magnitude' comes from the Latin 
\emph{magnus} (big, large) and \emph{magnitudo} (measure of bigness).
Here `measure' means \emph{unit}, which suggests that
magnitude is a quantified conception of some property: we speak of
magnitude when we can quantify some property and we can measure it,
no matter how indirectly. Think here of mass as
quantity of matter (Newton), momentum as quantity of motion (Huygens), volume as 
quantity of 3-dimensional space, acidity as quantity of acid in a solution (Arrhenius), 
biomass as quantity of matter produced in carbon, hydrogen and oxygen, electric current as 
quantity of electricity (Gilbert), and so forth.  
 
A general definition of magnitude is not around.  An appealing idea seems to define a 
magnitude as a \emph{quantified} or \emph{quantitative property}.
Measuring magnitude $\eulA$ of physical system $\sfS$ and obtaining value
$a$ would then show that $\sfS$ possesses a quantified property that we
could represent by: $\la\eulA,a\ra$. But this runs afoul against standard {\small QM},
which has taught us that measuring $\eulA$ definitely is \emph{not}
revealing a property possessed by $\sfS$ before the measurement. 
On the contrary, property $\la\eulA,a\ra$
gets ascribed to $\sfS$ \emph{just after} a measurement has ended and
the state of $\sfS$ collapes to an eigenstate that belongs to $a$, which then is an
eigenvalue of the representing operator $\widehat{A}$ acting on the Hilbert-space
$\calH$ associated with $\sfS$. 

Thus we take magnitude $\eulA$ as primitive and define a \emph{quantitative property} as
$\la\eulA,a\ra$, where $a\in\V(\eulA)\subseteq\R$, the set of values of $\eulA$,
or as $\la\eulA,a,\trm{u}(\eulA)\ra$ when magnitude $\eulA$ has a
\emph{unit}. If needed, $\V(\eulA)$ can include complex numbers, in which case 
$\V(\eulA)\subseteq\C$. 

A few examples ($\R^{+}$ contains $0$):
\beq \la\trm{mass},\;\R^{+},\;\trm{kilogram}\ra\;,\qquad
\la\trm{length},\;\R^{+},\;\trm{meter}\ra\;,\qquad
\la\trm{energy},\;\R,\;\trm{joule}\ra\;.\enq

We have now taken care of everything that is involved in the concept of 
measurement \eqref{Meas}. Next we present our explication of measurement.
\subsection{Main Dish}
{\up}Much of the labour we had to perform to arrive at an \emph{analysans} of our
\emph{analysandum}, that is, at a criterion for the core concept of measurement,
has already been performed in our analysis of a measurement apparatus. 
\begin{itemize}
{\item[]\textbf{Criterion for Measurement.}~~\emph{$p$ measures $\eulA$ of 
$\,\sfS$ by means of $\,\sfM$ and obtains $a\;$} iff\\
(1)~~$p$ is a person,\\
(2)~~$\eulA$ is a magnitude,\\
(3)~~$\sfS$ is a physical system,\\
(4)~~$\sfM$ is an $\eulA$-measurement apparatus,\\
(5)~~$a\in\V(\eulA)$ ($a$ is a value of $\eulA$),\\
(6)~~$p$ makes $\sfS$ and $\sfM$ physically interact 
$\eulA$-relevantly and this $\eulA$-relevant interaction results in
$\eulA$ having value $a$, which $\sfM$ registers or displays.}
\end{itemize}

Does this criterion cover all
measurements that have been, are and will be performed by anyone anywhere? 
I would be surprised if it did.
For example, how about measuring the length of the table by a 
tapeline? Is the result of $250\;$cm, the value of the length of the 
table, a result of a `lenght-relevant physical interaction between
table and tapeline'? Their interaction consists of no more than they
absorb some of each other's emitted electro-magnetic radiation$\ldots$
For another example, how about measuring time by a clock? When the clock is the measurement apparatus $\sfM$, what is the physical system $\sfS$? Perhaps also $\sfM$:
it measures the length of its worldline of spacetime, although that presupposes the
Theory of Relativity. But let's stop, and ask what a measurement interaction is. 
\begin{itemize}
{\item[]\textbf{Criterion for Measurement Interaction.}~~A physical interaction $I$ between two physical systems is a \emph{measurement interaction}~~iff~~there is a physical magnitude $\eulA$ such that at least one of the physical systems is an $\eulA$-measurement apparatus and $I$ is an $\eulA$-relevant physical interaction.}
\end{itemize}

This characterisation of measurement interaction is not entirely physico-ontological but 
partly empistemological, just as measurement is, due to our characterisation of
what an $\eulA$-relevant interaction is (see above). This is how it ought to be,
for to measure is to acquire knowledge. Measurement is also a species of knowledge
acquisition. Quantum-mechanical measurement theory provides more detailed mathematical
representations of measurement interactions.\endnote{See Suppes [2001], pp.~63--73, for
some general Measurement Theory; see Bush, Lahti and Mittelstaedt [1996] for physical
measurement theory. The concept of a measurement apparatus is not analysed but taken
for granted in both books, remarkably.} Back to the interpretation of {\small QM}.
\section{Building supra Building pon the Banks for the Livers by the Soangso}
{\up}The Prime Directive of Physics is that numbers calculated by using a physical
theory (or model or hypothesis or principle) 
should coincide with numbers measured that pertain to physical systems
the theory is supposed to be about.
Suppose there is a minimal set of postulates of {\small QM} in the sense that the
Prime Directive is obeyed: the postulates are just enough to
calculate measurement outcomes and their probability measures,
and these outcomes match what is being measured.
Call this: \emph{minimal {\small QM}} (soon to be characterised rigorously).

When the aim of physics is
\begin{itemize}
\setlength{\itemsep}{-1ex}
{\item to explain the (observed and unobserved)  phenomena, \emph{or}}
{\item to understand why things happen when they happen, \emph{or}}
{\item to find out what physical reality is like, what it is made of, 
what there is, what exists, what the properties and relations are of the actual beings, 
and how the actual beings behave and influence each other, \emph{or}}
{\item to reveal the structure of the universe as it is in and of itself, \emph{or}}
{\item any other aim that goes above and beyond merely calculating putative measurement outcomes,}
\end{itemize}
then, already then, minimal {\small QM} falls short of reaching the aim of science, 
for instance by telling us nothing about the fate of the cat and any other
physical system that is not measured. Minimal
{\small QM} leaves too many meaningful questions about physical
reality wide open. When minimal {\small QM} is a failure, must it not be refused entrance
to the body of scientific knowledge? Is the current presence of {\small QM}
in that body not a cyst which should be surgically removed? 

Nay nay, do not be afraid. I am not going to propose \emph{that}. 
The presence of minimal {\small QM} is wonderful,
provided we \emph{extend} it so as to approach the aim of physics more closely.
To provide an interpretation of {\small QM} is, we submit,
\emph{to add postulates to those of minimal {\small QM} so as to provide answers
to questions about physical reality that we deem meaningful and that pertain to
physical systems falling within the purview of minimal {\small QM}};  extending 
minimal {\small QM} may very well involve \emph{changing and usually extending its sparse 
vocabulary}. Van Fraassen:
\begin{quote}
{\small  Ideally, belief presupposes understanding. This is true even of the mere belief that a 
theory is true in certain respects only. Hence we come to the question of \emph{interpretation}: 
under what conditions is this theory true?  What does it say the world is like? 
These two questions are the same. \\
($\ldots$)\\
Suppose we agree that there can, in logical principle, be more than one adequate interpretation of 
a theory. Then it follows at once that interpretations go beyond the theory; the theory plus
interpretation is \emph{logically stronger} than the theory itself. For how could there be differences between views, all of which accept the theory, unless they vary in what they 
add to it?}\endnote{Fraassen [1991], p.~242--243. What Van Fraassen here calls
``the theory'' will be our minimal QM.}
\end{quote}
There may be hermeneutical activity in the wake of extending minimal {\small QM}
in the literal sense of the word, in that the meaning of certain expressions have to be adjusted to 
fit the intended extension of mininal {\small QM}, but the core interpretational activity
is to extend mininal {\small QM} by adding postulates, which implies (\emph{salute} Van
Fraassen) to provide {\small QM} with logically stronger truth-conditions (than the ones of minimal {\small QM}).

What is minimal {\small QM} precisely? Here follows an attempt to characterise it, 
call it $\minqm$. Let
$\calI(\R)$ be a Boolean subset algebra of closed intervals of the real line.
\begin{famlistdik}
{\item[\textbf{P0. Hilbert-Space Postulate (Von$\;$Neumann).}]
\emph{Associate some Hilbert-space $\calH$ to physical system $\sfS$, and a direct-product
Hilbert-space to a composite physical system with the factor Hilbert-spaces being
associated to the disjoint subsystems.}}
{\item[\textbf{P1. Evolution Postulate (Schr\"{o}dinger).}] 
\emph{Time is represented by the real continuum} ($\R$). \\
\emph{IF no measurements are performed in time-interval} $\Delta\in\calI(\R)$ 
\emph{on physical system $\sfS$}, \\
\emph{THEN at every moment in time $t\in\Delta$,
associate a Hilbert-vector} $|\psi(t)\ra\in\calH$ (\textbf{P0})
\emph{to $\sfS$ such that there is a connected Lie-group
of unitary operators acting in $\calH$ such that $|\psi(t)\ra=U(t)|\psi(0)\ra$, where 
$|\psi(0)\ra$ is associated to $\sfS$ at time $t=0$, and where $U(t)$ is a group member, such that $U(t+t')=U(t)U(t')$, for every $t,t'\in\Delta$.}}
{\item[\textbf{P2. Magnitude Postulate (Von$\;$Neumann).}] 
\emph{Represent physical magnitudes of interest by operators acting in $\calH$} (\textbf{P0}) \emph{that
have a spectral resolution. Restrict the domain of this resolution to $\calI(\R)$,
so that we consider only:
$\calI(\R)\to\calP(\calH)$, $\Delta\mapsto P^{B}(\Delta)$, where
$P^{B}(\Delta)$ is a projector from the Hilbert-lattice $\calP(\calH)$ that belongs to the spectral resolution of $B$}.}
{\item[\textbf{P3. Probability Postulate (Born).}] 
\emph{The probability for finding a value in interval $\Delta\in\calI(\R)$, at time $t$,  
upon measuring physical magnitude represented by operator $B$} (\textbf{P2})
\emph{when Hilbert-vector $|\psi(t)\ra\in\calH$ is associated to
$\sfS$ at time $t$} (\textbf{P0}, \textbf{P1}), \emph{equals the 
expectation-value of $P^{B}(\Delta)\in\calP(\calH)$;
in Redhead-notation}:
\beq\hspace*{-2.5em} \textrm{Pr}\big([B]^{|\psi(t)\ra}\in\Delta\big)\,=\,
\la\psi(t)|P^{B}(\Delta)|\psi(t)\ra \;.\label{Born}\enq}
\end{famlistdik}
For the sake of brevity, we have left out the Symmetrisation Postulate,
which is about composite systems
of similar particles (Bose-Einstein, Fermi-Dirac statistics).

Notice that $\minqm$ only speaks of \emph{physical systems, physical magnitudes and
probability distributions over measurement outcomes}. The theory $\minqm$ is
sufficiently strong to enjoy an extremely wide variety of confirmation.
We point out that physical magnitudes
can be identified with equivalence classes of measurement procedures, so
`physical magnitude' can be eliminated from the primitive physical vocabulary
(at the price of adding `measurement procedure').
Not a word in $\minqm$ about physical states,
physical properties and physical relations. No not one. \emph{Stricto sensu} $\minqm$
is a mathematical recipe to calculate probability distributions over measurement outcomes.
$\minqm$ says little if anything about physical reality outside the laboratory,
let alone about the microphysical world. This is unacceptable.

$\minqm$ does not include the notorious projection postulate (for a moderately precise statement, see below). Can $\minqm$, then, deal with repeated measurements? If
not, $\minqm$ may be an empirical failure. An adherent of $\minqm$ 

The game of physics. One group of people, the Experimenters, produce numbers by manipulating
various technical artifacts, and another group of people, the Theoreticians, think
of mathematical recipes that also produce numbers. The aim of the game
is that those numbers should match. The Experimenters usually begin and the Theoreticians
then must match whatever the Experimenters come up with. If the Theoreticians fail, they
lose and the Experimenters win; if the Theoreticians succeed, they win and the
Experimenters lose. Sometimes the Theoreticians begin and then the Experimenters
have to match. This is the game of physics, even the game of
science, in a nutshell I take it. But \emph{why} do we play this game? 
Out of boredom? For the helluvit? I say: \emph{No no no}. We play it because we want
the Theoreticians to win, because when they win repeatedly with the same theory,
that theory may be knowledge of physical reality, may provide explanations of the
phonemena that make us understand physical reality, and gathering such knowledge is
the epistemic aim of physics. Otherwise the
repeated success of the theory would be a miracle and we don't believe in miracles.

\emph{Standard}, or \emph{orthodox quantum mechanics} ($\stqm$) qualifies as an 
interpretation in the sense above of being an extension of $\minqm$, for it enriches the 
vocabulary of $\minqm$, strengthens some of its postulates and adds new postulates to it.

The language of $\stqm$ includes: physical properties and physical
state. The Hilbert-Space Postulate (\textbf{P0}) becomes the
\display{\blsq~\textbf{Pure State Postulate (Von$\;$Neumann).} 
\emph{Every possible pure physical
state of a physical system is mathematically represented by a normed
vector in some Hilbert-space, which we associate with the physical system.}}
(The `pure' alludes to a more general State Postulate encompassing also
mixed states, which are not mathematically represented by Hilbert-vectors.
We gloss over this.)
Also the Standard Property Postulate is added, as well as the controversial
\display{\blsq~\textbf{Projection Postulate (Dirac, Von$\,$Neumann).} \emph{IF
one performs a measurement of physical magnitude $B$ on a physical system,
when it has state $|\psi(t)\ra\in\calH$ at the moment $t\in\R$ of measurement,
AND one finds outcome in $b\in\Delta\in\calI(\R)$, with $\Delta$ the measurement
accuracy of measuring value $b\in\Delta$,\\
THEN immediately after the measurement outcome $b\in\Delta$
has been obtained, the post-measurement state of the physical system is represented by 
$P^{B}(\Delta)|\psi(t)\ra$.}}
\indent The Probability Postulate (\textbf{P3}) entails that the probability
of finding a measurement-outcome, when measuring physical magnitude $B$,
that does not lie in the spectrum of $B$ vanishes. Since it depends on one's
interpretation of probability of whether it follows that finding a measurement-outcome 
that is not in the spectrum of $B$ is impossible,
an explicit postulate is needed to exclude this. Here it comes.
\display{\blsq~\textbf{Spectrum Postulate (Schr\"{o}dinger, Von$\,$Neumann).}
\emph{All and only values from the spectrum of an operator that represents
a physical magnitude are its possible measurement-outcomes.}}
\indent So much for minimal $\minqm$ and its standard interpretation  (aka orthodox quantum mechanics: $\stqm$).
Let us turn for a moment to a few other interpretations. 
\section{The Ineluctible Morality of the Intelligible}
{\up}Bohr's Copenhagen Interpretation has long been, and perhaps still is,
the interpretation most physicists adhere to. It adds the following postulates to $\minqm$,
resulting in, say, $\copqm$.
\display{\blsq~\textbf{Quantum Postulate (Bohr).} \emph{Every quantum phenomenon is
indivisible; disconnected considerations of its parts are inappropriate,
because the interaction between object-system and preparation and registration
apparatus is not eliminable due to Planck's constant $(h>0)$}.}
By the \emph{quantum phenomenon}, Bohr means the whole of the preparation apparatus,
which one uses to prepare the object-system in a particular physical state,
the registration apparatus one uses to measure some physical magnitude,
and of course the physical system that is being subjected to preparation and
measurement, the \emph{object-system}.
In classical physics one can appropriately consider parts, without mentioning
other parts or the whole. In $\copqm$ the Quantum Postulate rules, which
is however limited by the
\display{\blsq~\textbf{Buffer Postulate (Bohr).} \emph{The literal
description of preparation and registration apparatus, and of the measurement
outcomes, is given in the language of classical physics;
the \emph{Deutung} of the object-systems proceeds by means of
mathematical concepts.}}
Finally there is the
\display{\blsq~\textbf{Complementarity Postulate (Bohr).} \emph{The quantum phenomenon,
specifically the experimental arrangement of the pieces of measurement apparatus
(preparation and registration apparatus), determines which classical concepts are
applicable. There are pairs of classical concepts, like wave/particle,
kinematics/dynamics, space-time/causality, that are never
jointly applicable in a single experimental arrangement but only in mutually exclusive
experimental arrangements and in this way provide an exhaustive description
of the object-system. Such pairs are called \emph{complementary}. They are
however jointly applicable in so far as the relevant Indeterminacy Inequality
permits.}}
According to Bohr, the language of classical physics is indispensable
for {\small QM}. Bohr viewed this language as a \emph{refinement} of Nalasweknowit:
material objects in space, that persist over time and whose
properties change over time as a result of causal processes. The `classical
language' is unambiguous and accurate, so that the objectivity
of {\small QM} is guaranteed.

By \emph{classical science} in general, Bohr meant scientific inquiry
where the role of the scientist, the subject and his thought and talk, 
can be ignored, thus resulting in a
subject-independent hence objective description or explanation of a part of
reality that falls within the relevant scope of scientific inquiry.
Classical physics, usually by definition the whole of physics accepted in
the year 1900, qualifies as `classical' in Bohr's sense.  
Classical physics is needed to guarantee the objectivity
of {\small QM}.

All modal interpretations obviously qualify as interpretations 
of {\small QM} and
are much more modest in their extensions of the vocabulary of $\minqm$ than
Bohr \emph{cs}. One modal interpretation rejects the projection postulate of {\small QM},
rejects measurement as a primitive concept in the vocabulary,
makes the Evolution Postulate (\textbf{P1}) hold unconditionally, and
replaces the standard property postulate with the Sufficiency
Property Postulate and the
\display{\blsq~\textbf{BiModal Property Postulate (Dieks-Vermaas).}
\emph{The subsystems of a composite system have one of the quantitative 
properties $\la B,b\ra$, such that the basis of the 
Schmidt bi-orthogonal decomposition of the state of the composite system 
is the eigenbasis of $B$,
with probability as in the Probability Postulate} (\textbf{P3}).}

The Everett interpretation also qualifies as an interpretation of {\small QM}
because it changes the vocabulary of $\minqm$ (adding the concept of a \emph{branch}, or a \emph{perspective}, or a \emph{world},  and deleting the concept of measurement
as primitive), adds a branching postulate:
\display{\blsq~\textbf{Branching Postulate (Everett).}
\emph{Consider a particular basis of the Hilbert-space $\calH$, associated with any physical
system $\sfS$, and expand its physical state $|\psi\ra\in\calH$ (State Postulate) 
in this basis, say
$|\phi_{j}\ra\in\calH$, for $j=1,2,\ldots$ dim$(\calH)$.
Then relative to branch $j$, $\sfS$ has physical property $\la B,b_{j}\ra$, where
$B|\phi_{j}\ra=b|\phi_{j}\ra$.}}

A solution of the problem \emph{which} basis to consider is nowadays sought by
an appeal to `decoherence', which is the generic phenomenon that when a physical
system $\sfS$ is in a physical environment (radiation, heat bath, air), 
the state becomes diagonal in some particular
basis, the `decoherence basis'. Often this basis corresponds to the physical
magnitude energy or position, and it is this basis, `preferred' so to speak
by Mother Nature, that is then considered in the Everett Postulate above,
notably by Oxonian Everettians. They thus have physical reasons to attach 
ontological significance to the terms of $\psi$ 
when expanded in one basis rather than an infinitude of other bases ---
perhaps even excellent physical reasons ---,
but that does not mean that they do adere ontological significance to
\emph{these} terms, and that means that $\evqm$ goes above and 
beyond $\minqm$, --- and, of course, differs from $\stqm$.

Even Bohmian Quantum Mechanics  ($\bqm$) qualifies. 
$\bqm$ adopts the Hilbert-space of complex wave-functions on configuration
space. For the sake of simplicity, we consider $2$ spinless particles in $3$-dimensional 
space, having mass $m_{1}$ and $m_{2}$. The wave-function of the composite system is: 
$L^{2}(\R^{3})\otimes L^{2}(\R^{3})\,\simeq\,L^{2}(\R^{6})$.
\display{\blsq~\textbf{Bohmian State Postulate.}
\emph{The \emph{state} of this $2$-particle system is represented by:
$\la\psi,\bfQ\ra$, where $\psi: t\mapsto\psi(t)\in L^{2}(\R^{6})$} (\textbf{P0. 
Hilbert-Space Postulate}) \emph{and} $\bfQ:t\mapsto\bfQ(t)\in\R^{6}$ (\textbf{Position Postulate}: see below.)}
Just as in $\minqm$, $\psi$ is postulated to obey the Schr\"{o}dinger equation.
Vector $\bfQ(t)$ consists of $6$ components, and can be written as
$\la\bfQ_{1}(t),\bfQ_{2}(t)\ra$, where $\bfQ_{1}(t)$, $\bfQ_{2}(t)\in\R^{3}$.
Vector $\bfQ_{1}(t)$ represents the position of $\bfp_{1}$ at time $t$ and
similarly $\bfQ_{2}(t)$. Like in classical mechanics but unlike in $\stqm$, in
$\bqm$ every particle always has a position.
Bohmians `complete' {\small QM} by adding $\bfQ$ to $\psi$.
\display{\blsq~\textbf{Position Postulate.}
\emph{The positions of the particles are determined by $\psi$ via the 
Guiding Equation, which is for particle~1}:
\beq m_{1}\frac{d\bfQ_{1}(t)}{dt}\,=\,\hslash\,
\mrm{Im}\left(\frac{\nabla_{\!1}\,\psi(\bfq_{1},\bfq_{2},t)}
{\psi(\bfq_{1},\bfq_{2},t)}\right)_{\bfq_{1}=\bfQ_{1}(t)}\;,
\label{GuidingEq}\enq
\emph{where $\nabla_{\!1}$ is the gradient, with respect to $\bfq_{1}$, and}
Im$(z)\in\R$ \emph{is the imaginary part of $z\in\C$. Similarly for particle~2.}}
One should not confuse $t\mapsto\bfQ_{1}(t)$ with $\bfq_{1}$: the afore-mentioned 
describes the path of particle~1 in $3$-dimensional space, whilst the
last-mentioned is a physically uninterpreted variable of $\psi$.

The left-hand-side of the Guiding Equation \eqref{GuidingEq}
is a time-derivative of the position of particle~1, which is the definition of its
velocity:
\beq \textbf{v}^{\psi}_{1}(t)\,=\,\frac{d\bfQ_{1}(t)}{dt}\;, \enq
where the superscript `$\psi$' is there to emphasise that the velocity is determined
by $\psi$, via eq.\ \eqref{GuidingEq}, which pertains to the composite system.
There is further an {\blsq}~\tbf{Equilibrium Postulate}, which posits the Born-measure for
position probabilities. From this and an elaborate story that reduces all
measurements to position measurements, the Probability Postulate follows.

\emph{Legenda table below.} All theories entail the postulates of $\minqm$,
which are therefore omitted; only the additional postulates are mentioned. By
`{\small $1/2$}'  is meant the Sufficiency Property Postulate. {\blsq}~\tbf{Categorical
Evolution Postulate}: always unitary evolution over time, whether measurements
are performed or not.  \\
\begin{center}
\begin{tabular}{l |c |c |c | c| c}
&$\stqm$&{\small BiModQM}&{\small CopQM}&{\small EvQM}&{\small BQM}\\ \hline 
\tbf{St.\ Prop.\ Post.} & + & {\small $1/2$} & {\small $1/2$}&{\small $1/2$}&--\\ \hline
\tbf{Pure State Post.} & + & + & + & + &+\\ \hline
\tbf{Projection Post.} & + & -- & +/-- & -- & -- \\ \hline
\tbf{Spectrum Post.} & + & + & + & + & + \\ \hline
\tbf{Categ.\ Evol.\ Post.} & + & + & +/-- & + & + \\ \hline
\tbf{Quantum Post.} & -- & -- & + & -- & -- \\ \hline
\tbf{Buffer  Post.} & -- & -- & + & -- & -- \\ \hline
\tbf{Compl.\ Post.} & -- & -- & + & -- & -- \\ \hline
\tbf{BiMod.\ Prop.\ Post.} & -- & + & -- & -- & -- \\ \hline
\tbf{Branching Post.} & -- & -- & -- & + & -- \\ \hline
\tbf{Bohm.\ State Post.} & -- & -- & -- & -- & + \\ \hline
\tbf{Position Post.} & -- & -- & -- & -- & + \\ \hline
\tbf{Equilibr.\ Post.} & -- & -- & -- & -- & + \\ \hline
\end{tabular}
\end{center}
\mbox{}\\
\section{True Inwardness of Reality}
\label{SectReality}
{\up}What we have not done is to expound yet another interpretation of {\small QM}, to defend
one or to criticise one. What we have done is something more modest. We have
expounded what it means \emph{to interpret} {\small QM} and it means, in a 
nutshell, this: to extend $\minqm$ by adding postulates and enriching
the vocabulary. This is achieved by proceeding as follows. 
\begin{famlistdik}
{\item[\ding{182}] List the concepts that the interpretation employs
\emph{in addition to} those of minimal {\small QM} ($\minqm$), which are: physical system, physical subsystem, physical magnitude, probability, measurement; explain these additional
concepts.}
{\item[\ding{183}] Mention whether the physical concepts of  $\minqm$ change in the
new interpretation, i.e.\ whether the meaning of the words expressing them differs
in the new interpretation when these words are already employed in $\minqm$; 
explain these differences.}
{\item[\ding{184}] List the postulates that the new interpretation 
adds to those of $\minqm$; if postulates of $\minqm$ are not among those of the
new interpretation, show that these postulates of $\minqm$ become theorems in
the new interpretation.}
{\item[\ding{185}] Mention whether some (or all) of the postulates 
of $\minqm$ change in the new interpretation; explain these changes.}
{\item[\ding{186}] List the questions that minimal $\minqm$ does not answer,
or the problems that $\minqm$ does not solve, and show how the new interpretation
of {\small QM} answers (some of) them and solves (some of) them, respectively.}
\end{famlistdik}

The above list ought to be the to-do list for every interpreter of {\small QM}.

Thus the interpretation of {\small QM} turns out to be \emph{not the
same as} how `to interpret' is generally interpreted in philosophy, which is: \emph{to assign
meaning to}. Depending on the interpretation under consideration, there is more or less
of interpretation in the last-mentioned sense going on; but the thesis that this is \emph{all} 
that is going on in the discourse on the interpretation of {\small QM} is like saying that 
arranging the table is all that is going on in the preparation of a dinner.

Is the interpreation of {\small QM}, then, perhaps a special
case of hermeneutics as we have come to know it in continental philosophy, 
where we think of the likes of Schleiermacher, Dilthey, Heidegger, Gadamer and Derrida? 
Is the discourse on the interpretation of {\small QM} an hermeneutical discourse in
his sense, i.e.\ is there such a thing as \emph{quantum hermeneutics}?  
Let us briefly take a closer look at hermeneutics in philosophy.

The word `hermeneutics'  comes from the Greek word for interpretation or translation
($\varepsilon\rho\mu\eta\nu\varepsilon\upsilon\omega$), 
which derives from the name of the Greek mythological figure Hermes, 
who deciphered messages of the gods and communicated them to human mortals.
Aristotle introduced hermeneutics in philosophy in his \emph{De Interpretatione}, by
distinghuishing the symbols or signs (\emph{symbola})
from the affect they have on our minds (\emph{pathemata}) as well as from the
entities they represent (\emph{pragmata}), of which the mental affections 
are representations (\emph{homoiomata}). Hermeneutics in philosophy is the study of written 
texts in context, in order to understand the text better, notably to come to know what the
text expresses, to which the text provides access. The study of sacred texts in Talmudic, Vedic,  
Biblical and Apostolic traditions belong to 
\emph{theological} or \emph{religious} hermeneutics; they have one leg in mythology (Hermes) 
and the other one in philosophy (Aristotle).
The context of the text is usually taken to be the historical context in which the
text is produced (Dilthey), in order to understand the views and intentions of the author 
(Schleiermacher) or to understand what the text itself expresses (Dilthey), where
`understanding' has to be understood in the sense of Droysen's \emph{verstehen} rather 
than \emph{erkl\"{a}ren}.\endnote{Wright [1971], p.~5.} Heidegger gave birth to 
\emph{existential hermeneutics},
an endeavour to understand human existence, \emph{Dasein}, directly, without mediation
by text and language generally. Inquiry into written text in context, call it
\emph{textual hermeneutics}, is something else: indirect and further removed from
life as we live and experience it.  Existential hermeneutics was further developed by 
Heidegger's pupil Gadamer [1960], who further delved into individual human experience, mediated by language however, in particular by spoken language in conversation.

The \emph{hermeneutic circle}, an idea introduced by Heidegger, 
has various manifestations. One is that in order to understand parts of a text, one
needs to understand the text as a whole, and in order to understand the whole text,
one needs to understand its parts. The process of interpretation, leading
to an ever increasing understanding, thus proceeds in a circle of reading and re-reading. 
One understands the postulates of {\small QM} better after one has understood the whole of
{\small QM}, and one understands the whole of {\small QM} better after one
has understood the postulates. This is however not what is going on
in the discourse of the interpretation of {\small QM}. Another manifestation of the
hermeneutic circle is the reciprocity between text and context. But inquiry into
the historical context of the advent of {\small QM}, and into what the effect 
of the historical context on the content of quantum-mechanical texts 
has been belong to the discourse of the history of {\small QM},
not to the interpretation of {\small QM}. So this second manifestation of the
hermeneutical circle also is definitely not what is going on in the discourse of the interpretation
of {\small QM}. 

Derrida took a turn in textual hermeneutics by considering
only \emph{other texts} as the context of a text, leading to his notorious
assertion ``There is nothing outside the text.''\endnote{Derrida [1967], pp.~158--159. 
Derrida was parenthetically heavy influenced by \emph{Finnegans Wake},  see Derrida 
[1984].} ``There is nothing outside context'', expresses the same, Derrida later
explained.\endnote{Derrida [1988],  p.~148.}  Notice that we only have access to the past, 
to the factual historical context in which a text is written, by means of other texts ---
and occasionally images and artefacts.
Use of words in other texts resonate in the text under consideration, and their use in 
the text under consideration resonate back in all other texts. This seems yet another 
manifestation of the hermeneutic circle. But, again, this hardly helps to capture what 
is going on in the interpretation of {\small QM}. 

Parenthetically, in \emph{Finnegans Wake} Hermeneutic Circles are Everywhere 
({\small HCE}).

We tentatively conclude that there is no such thing as `quantum hermeneutics'.\endnote{In the 
sense of hermeneutics as understood in the philosophical tradition sketched above. In another,
literal sense, `quantum hermeneutics' just means `quantum interpretation'.}
This conclusion savours an \emph{a priori} possible and perhaps promising
connexion between the discourse of (i) philosophy of physics and of (ii)
hermeneutics in philosophy --- and philosophy of language we submit. 
Interpreting {\small QM} is not merely `a matter of semantics' or penetrating
deeper into quantum-mechanical texts and their context. 
What is at stake in the discourse of the interpretation of {\small QM} 
is \emph{how} and \emph{what} microphysical reality is,  \emph{how to understand} 
microphysical reality --- if it is understandable by us at all ---, granted that {\small QM}
provides us with the best basis to answer these questions. What is at stake here are
answers to all sorts of questions concerning microphysical reality, the world of the tiny and the brief, and to physical reality generally.
Hopefully the answers to these questions jointly provide some coherent
understanding of physical reality. Finding answers becomes a matter of finding the
right additional postulates to extend $\minqm$, rather than just keeping the 
postulates fixed and re-interpreting expressions occurring in them. 
Novel concepts, alien to Nalasweknowit, not in use and nowhere expressed in other 
texts, may very well have to be constructed for this purpose. Steps \ding{183}, \ding{184} and \ding{185} are supposed to involve precisely radical conceptual change.

Hence in one of the most successful 
areas of natural science, quantum physics, an interpretational inquiry was
launched by theoretical physicists in the 1920ies; later philosophers 
joined in, with a vengeance. A mainstream interpretation was settled, of Copenhagen design. 
But it did not last. Copenhagen {\small QM} has left too many questions unanswered. 
Schr\"{o}dinger complained that the interpretational problems of {\small QM}
were shelved, not solved. In his Nobel Lecture of 1969,
Murray Gell-Mann notoriously declared that an entire generation
of physicists was brainwashed into believing that the interpretation problems of {\small QM}
were solved, by the Great Dane.
To interpret {\small QM} is to extend minimal $\minqm$ and its vocabulary, 
which permits the expression of more concepts than the language of $\minqm$ permits. 
Since forging novel concepts is a philosophical activity \emph{par excellance},
a philosopical activity is required to aid physics to achieve its aims.

We want to unveilop the theory of the building blocks of matter and their
interactions. We need to destructify the obscuritads it bangcreated, in order to sunshine 
physical reality by a syntasm of ison and remagination.  For
we want to understand. We need to understand. We shall understand.
In our stream of consciousness, from swerve of shore to bend of bay, we
want to float and fly, transpicuously and persparantly, and not to drown and 
die.

The final word is to B.C.\ van Fraassen, with an empiricist twist at the end:
\begin{quote}{\small
Why then be interested in interpretation at all? If we are not interested in the 
metaphysical question of what the world is really like, what need is there to look into these 
issues?\\[1em]
Well, we should still be interested in the question of how the world could be the way quantum 
mechanics --- in its metaphysical vagueness but empirical audacity --- says it is. 
That is the real question of \emph{understanding}. To \emph{understand} a scientific theory, 
we need to see how the world could be the way that the theory says 
it is. An \emph{interpretation} tells us that. The answer is not unique, because the question 
`How could the world be the way the theory says it is?' is not the sort of question to call for 
a unique answer. Faith in the actual truth of a good answer, so interpreted, is neither required by 
\emph{understanding}, nor does it help.\endnote{Fraassen [1991], p.~337.}}
\end{quote}
\section{Patrick Colonel Suppes}
{\up}Permit me to admit that it feels awkward to contribute to a Festschrift for
Patrick Colonel Suppes while not having attended the actual celebration conference 
at Stanford University that ook place in March 2012.
Perhaps as awkward as contributing a paper about a subject about which Patrick
Suppes has been suspiciously silent in all of his writings about {\small QM}.
Does he fancy the Copenhagen Interpretation? Does he prefer some modal interpretation?
Is he an Everettian? No favourites at all? Lame agnosticism? Brute rejection of
the very issue of the interpretation of {\small QM} as a pseudo-issue? 
Is he aware of the obscuritads that circumveilop us? 
 
Suppes has not been entirely silent about {\small QM}. Everything he has written about
{\small QM} is about or connected to \emph{probability} (available
at his website of Stanford University, spanning \emph{seven} decades).\endnote{Suppes [1963],
[1965].}
Concerning the issue of  the \emph{the interpretation of probability}, however, 
in and outside {\small QM}, Suppes has also been suspiciously silent about where his sympathy lies, his insistence that the differences in interpretation of probability ought
to be characterised mathematically notwithstanding.\endnote{Suppes [2001], Chapter~5.}
Is he a Bayesian? Or a frequentist? Does he believe in propensities?

Let this paper then be the trigger for an old wise scientific philosopher 
to speak, at last, his mind on these exciting philosophical issues $\ldots$\\[3em]
{\small \emph{Acknowledgments.} Thanks to Gijs Leegwater, Geurt Sengers and Stefan Wintein (Erasmus
University Rotterdam), and an anonymous Referee for comments.  \\ \\
\emph{Affiliations.} Faculty of Philosophy, Erasmus University Rotterdam,
Burg.\ Oudlaan 50, H5--16, 3062 PA Rotterdam, E-mail: f.a.muller@fwb.eur.nl; and:
Institute for the History and Foundations of Science,
Dept.\ of Physics \& Astronomy, Utrecht University,
Budapestlaan 6, IGG--3.08, 3584 CD Utrecht, E-mail: f.a.muller@uu.nl; The Netherlands}
\newline
{\addcontentsline{toc}{section}{Notes}%
\theendnotes}
\mbox{}\\[3em]
\noindent\tbf{\large References}
\addcontentsline{toc}{section}{References}%
{\small \begin{refs}
{\item Bush, P., Lahti, P.K., Mittelstaedt, P. [1996], \emph{The Quantum Theory of Measurement} (2nd.\ Rev.\ Ed.), Berlin: Springer-Verlag.}
{\item Derrida, J. [1976]. \emph{Of Grammatology}, translated from the French original
of 1967 by G.C.\  Spivak, Baltimore \& London: Johns Hopkins University Press.}
{\item Derrida, J. [1984]. `Two words for Joyce', in: \emph{Post-structuralist Joyce: Essays from the French}, D.\ Attridge and D.\ Ferrer (eds.), Cambridge: Cambridge University Press, pp.~145--160.}
{\item Derrida, J. [1988]. \emph{Limited Inc.},  Evanston: Northwestern University Press.}
{\item Dirac, P.A.M. [1928]. \emph{The Principles of Quantum Mechanics}, Cambridge:
Cambridge University Press.}
{\item Dummett, M.A.E. [1991]. \emph{The Logical Basis of Metaphysics}, Cambridge, Massachusetts: Harvard University Press.}
{\item Fraassen, B.C.\ van. [1991]. \emph{Quantum Mechanics: An Empiricist View},
Oxford: Clarendon Press.}
{\item Gadamer, H.-G. [1960]. \emph{Wahrheit und Methode: Grundz\"{u}ge einer philosophischen Hermeneutik}, T\"{u}bingen: Mohr.}
{\item Joyce, J. [1939].  \emph{Finnegans Wake}, London: Faber and Faber.}
{\item Muller, F.A. [2005]. `The Deep Black Sea: Observability and Modalify Afloat',
\emph{British Journal for the Philosophy of Science} \textbf{56}, pp.~61--99.}
{\item Neumann, J.\ von. [1932]. \emph{Mathematische Grundlagen der Quanten Mechanik},
Berlin: Springer-Verlag.}
{\item Saunders, S.W. [1996a].  `Time, Quantum  Mechanics,  and  Tense', \emph{Synthese} \tbf{107}, pp.~19--53.}
{\item Saunders, S.W. [1996b].  `Naturalizing Metaphysics',  \emph{The Monist} \tbf{80}, pp.~44--69.}
{\item Suppes, P.C. [1963]. `The role of probability in quantum mechanics', in: \emph{Philosophy of Science: the Delaware Seminar}, Volume~2, B.\ Baumrin (ed.), New York: Wiley \& Sons, pp.~319--337.}
{\item Suppes, P.C. [1965]. `Probability Concepts in Quantum Mechanics', \emph{Philosophy of Science} \textbf{28.4}, pp.~378--389.}
{\item Suppes, P.C. [2001]. \emph{Representation and Invariance of Scientific Structures}, Chicago: University of Chicago Press.}
{\item Wallace, D. [2012]. \emph{The Emergent Multiverse.\ Quantum Theory according to the
Everett Interpretation}, Oxford: Oxford University Press.}
{\item Wright, G.H.\ von [1971]. \emph{Understanding and Explanation}, 
London: Routledge \& Kegan Paul.}
\end{refs} }
\end{document}